\documentstyle[preprint,aps]{revtex}
\begin{document}
\draft
\title{Chiral Bosons and Improper Constraints\cite{byline}}
\author{F. P. Devecchi and H. O. Girotti}
\address{ Instituto de F\'{\i}sica,
Universidade Federal do Rio Grande do Sul \\ Caixa Postal 15051, 91501-970  -
Porto Alegre, RS, Brazil.}
\date{September 1993}
\maketitle
\begin{abstract}
We argue that a consistent quantization of the Floreanini-Jackiw model, as a
constrained system, should start by recognizing the improper nature of the
constraints. Then, each boundary
condition defines a problem which must be treated separately. The model is
settled on a compact domain which allows for a discrete formulation of the
dynamics; thus, avoiding the mixing of local with collective coordinates. For
periodic boundary conditions the model turns out to
be a gauge theory whose gauge invariant sector contains only chiral
excitations. For
antiperiodic boundary conditions, the model is a second-class theory where the
excitations are also chiral. In both cases, the equal--time algebra of the
quantum energy-momentum densities is a Virasoro algebra. The Poincar\'e
symmetry holds for the finite as well as for the infinite domain.
\end{abstract}
\narrowtext

Self dual fields, also known as chiral bosons, are of interest due to its
relevance in the heterotic string \cite{Gr} and in the quantum Hall
effect \cite{We}. The quantization of these objects is
beset with difficulties. Indeed, the Lorentz invariant model
in Refs.\cite{Sr,Mu}, based on the idea that the chiral condition could be
implemented through a linear constraint, does not exhibit physical
excitations \cite{Gi1}.  On the other hand, the canonical quantization of
Siegel's Lagrangian \cite{Si} is afflicted by an anomaly
which is to be eliminated by the addition of a Wess-Zumino term.
It turns out, then,  that the resulting theory does not describe
pure chiral bosons but rather their coupling to gravity \cite{Im}.

Of particular interest is the (1+1)--dimensional model put forward by
Floreanini and Jackiw (FJ)\cite{Fl,F1}, whose dynamics is described by the
noncovariant Lagrangian density
\begin{equation}
{\cal L} = \frac{1}{2} (\partial_{0} \phi) (\partial_{1} \phi) -
\frac{1}{2} (\partial_{1} \phi) (\partial_{1} \phi) , \label{1}
\end{equation}
where  $x \equiv (x^{0},x^{1})$ and $\phi \equiv \phi (x)$ is a real scalar
field whose canonically conjugate momentum will be denoted as $ \pi (x) $.
This model was quantized through the
Dirac-bracket procedure \cite{Di} by one of us (HOG) in collaboration with
M. E. V. Costa \cite{Co,F2}. It possesses a non-denumerable set of
constraints,
\begin{equation}
\gamma(x^{0},x^{1}) \equiv \pi(x^{0},x^{1}) - \frac{1}{2} \partial_{1}
\phi(x^{0},x^{1}) \approx 0, \label{2}
\end{equation}
while the canonical Hamiltonian is
\begin{equation}
H_{c} = \frac{1}{2} \int_{-\infty}^{+\infty} dx^{1} (\partial_{1}\phi)
(\partial_{1}\phi). \label{3}
\end{equation}
One readily verifies that the Poisson brackets of the constraints define a
matrix $ Q $,
\begin{equation}
Q(x^{0};x^{1},y^{1}) \equiv [\gamma(x^{0},x^{1}), \gamma(x^{0}, y^{1})]
= - \partial_{x^{1}} \delta (x^{1} - y^{1}), \label{5}
\end{equation}
whose inverse is not unique,
\begin{equation}
Q^{-1}(x^{0};x^{1},y^{1}) = -\frac{1}{2} \epsilon (x^{1} - y^{1}) +
\zeta (x^{0}). \label{7}
\end{equation}
In Ref. \cite{Co} the arbitrary function
$\zeta $ was set to zero. By abstracting the equal-time commutators from
the corresponding Dirac-brackets (Dirac-bracket quantization procedure) the
following equal-time commutation relations were, then, found\cite{Co}
\begin{mathletters}
\label{8}
\begin{eqnarray}
&&[\hat{\phi } (x^{1})\,,\, \hat{\phi } (y^{1})]  =  -\frac{i}{2} \epsilon
 (x^{1} - y^{1}),\label{mlett:1}  \\
&&[\hat{\phi } (x^{1})\,,\, \hat{\pi } (y^{1})]  =  \frac{i}{2}
 \delta (x^{1} - y^{1}), \label{mlett:2} \\
&&[\hat{\pi } (x^{1})\,,\, \hat{\pi } (y^{1})]  =  \frac{i}{4}
 \partial_{x^{1}} \delta (x^{1} - y^{1}), \label{mlett:3}
\end{eqnarray}
\end{mathletters}
where $\epsilon (x^{1})$ is the sign function and the carets denote operators.
One can check that the chiral field configuration
\begin{equation}
\label{9}
\hat{\phi} (x^{0},x^{1}) = \frac{1}{\sqrt{2\pi }} \int _{0}^{\infty} dk^{1}
\frac{1}{\sqrt{k^{1}}} \left[ \hat{\Lambda}
(k^{1})\,e^{-ik^{1}(x^{0}+x^{1})}\,\,+\,\,\hat{\Lambda}^{\dagger}
(k^{1})\,e^{ik^{1}(x^{0}+x^{1})} \right],
\end{equation}
with
\begin{equation}
\label{10}
\mbox [\hat{\Lambda}(k^{1}),\hat{\Lambda}(k^{\prime \, 1})]=
[\hat{\Lambda}^{\dagger}(k^{1}),
\hat{\Lambda}^{\dagger}(k^{\prime \,1})]=0;\,\,\,
[\hat{\Lambda}(k^{1}),\hat{\Lambda}^{\dagger}(k^{\prime \,
1})]=\delta(k^{1}- k^{\prime \, 1}),
\end{equation}
solves the Heisenberg equation of motion arising from  $ \hat{H_{c}}$ and the
equal-time commutation relations (\ref{8}).

The presence of the arbitrary function $\zeta$ in (\ref{7}) indicates that the
solution (\ref{9}) is not unique \cite{Wo} and, therefore, cast doubts on
whether all physical excitations of the FJ model are in fact chiral. To
elucidate this point is the main purpose of this work.

We start by recognizing that the constraints $\gamma$ are improper
\cite{Bg,St,Su}. Indeed, any function $\eta (x^{1})$ in the space
dual to the space of constraints $\gamma (x^{1})$ must verify
\begin{equation}
\label{11}
\delta \gamma [\eta] = \int_{- \infty}^{+ \infty} dx^{1} \left( \alpha
(x^{1})\,\delta \phi (x^{1})\,\,+\,\,\beta (x^{1})\,\delta \pi(x^{1}) \right),
\end{equation}
where
\begin{equation}
\label{12}
\gamma [\eta]\,\equiv \, \int_{- \infty}^{+\infty} dx^{1} \eta (x^{1}) \gamma
(x^{1}),
\end{equation}
$\alpha (x^{1}) \equiv \delta \gamma [\eta] / \delta \phi (x^{1})$ and
$\beta (x^{1}) \equiv \delta \gamma [\eta] / \delta \pi (x^{1})$.
{}From (\ref{2}) and (\ref{12}) follows that
\begin{eqnarray}
\label{14}
\delta \gamma [\eta ]& = & \int_{-\infty}^{+\infty}dx^{1} \left( \eta (x^{1})
\delta \pi (x^{1})\,\,+\,\,\frac{1}{2} \partial_{1} \eta (x^{1}) \delta \phi
(x^{1}) \right) \nonumber \\ & - & \frac{1}{2} \left (\eta (\infty ) \,\delta
\phi (\infty )\,\,-\,\,\eta (- \infty ) \,\delta \phi (- \infty ) \right).
\end{eqnarray}
The presence of the surface term in (\ref{14}) confirms the improper nature
of the constraints (\ref{2}). In order for the total Hamiltonian \cite{Di}
to be a proper generator of time transformations, one must require the
vanishing of this surface term.
Hence, the construction of the dual space depends on the boundary conditions of
$ \delta \phi (x^{1}) $. For instance, $ \delta \phi (\infty)=
\delta \phi (- \infty)$ demands $\eta (\infty)=\eta (- \infty)$,
while $\delta \phi (\infty)=- \delta \phi (- \infty)$ demands
$\eta (\infty)=- \eta (- \infty)$.

Thus, each set of histories $ \{ \phi (x) \}$ verifying a certain boundary
condition defines a problem which must be treated separately. In
this work the
FJ model is settled on a compact domain ($ -R \leq x^{1} \leq +R $) and then
quantized under periodic and antiperiodic boundary conditions. As
we shall see, to each boundary condition corresponds a different constraint
structure; in particular, only periodic boundary conditions allow for
first-class constraints. The quantum energy-momentum densities are constructed
and their equal--time algebra is investigated. A set of Poincar\'e charges is
built and the limit $R\rightarrow \infty $ is analyzed in both cases.

Let $ \{ \phi (x^{0}, x^{1})\,\, \vert \,\, \phi (x^{0}, + R)  =
\phi (x^{0}, - R) \}$ be the set of periodic histories and let
$\phi(x^{0}\,,\,x^{1}) $ be any history in this set.
One readily verifies that, in this case, the action
\begin{equation}
S[\phi ]\,\,=\,\,\int_{-R}^{+R} dx^{1} \left[ \frac{1}{2} (\partial_{0} \phi)
(\partial_{1} \phi) -
\frac{1}{2} (\partial_{1} \phi) (\partial_{1} \phi) \right] \label{20}
\end{equation}
is invariant under the transformation
\begin{equation}
\phi (x^{0},x^{1}) \longrightarrow \phi (x^{0},x^{1})\,+\,f(x^{0}), \label{21}
\end{equation}
which preserves the boundary conditions \cite{Bg}. This transformation,
which is neither local nor global, must be generated by a first--class
constraint. Such generator,  $\Gamma (x^{0})$, is an infinite combination of
the constraints (\ref{2}) and reads
\begin{equation}
\label{22}
\Gamma (x^{0})\,\,=\,\,\int_{-R}^{+R} dx^{1} \pi (x^{0},x^{1}) \approx 0.
\end{equation}
Of course, the invariance of the action under the transformation (\ref{21}) is
responsible for the lack of uniqueness present in (\ref{7}). This ``gauge
freedom'' can not be fixed by means of local gauge conditions. Alternatively,
one may try fixing by means of an integrated condition like
\begin{equation}
\chi (x^{0})\,\,=\,\,\int_{-R}^{+R} dx^{1} \phi(x^{0},x^{1}) \approx 0.
\label{23}
\end{equation}
When quantizing the theory along these lines one faces the problem of computing
Dirac brackets in which local field variables are mixed with
collective ones\cite{Sri}.

To free ourselves from the above drawbacks, we shall quantize the FJ model by
taking advantage of the compactness of the domain. This, together with the
the boundary conditions under analysis, allows for a discrete formulation of
the theory in terms of the real Fourier coefficients $a_{0}(x^{0}),
a_{n}(x^{0})$ and $ b_{n}(x^{0})$ entering in the decomposition of the real
field $\phi$\cite{F3},
\begin{eqnarray}
\label{25}
\phi (x^{0},x^{1})\,=\,\frac{1}{2R} a_{0}(x^{0})\,\, + \,\,\frac{1}{2R}
\sum_{n>0}\left\{ \left[a_{n}(x^{0})\,\, + \,\,i\,b_{n}(x^{0}) \right]\,
e^{\frac{in\pi } {R} x^{1}} \right. \nonumber \\
\left.  +  \left[a_{n}(x^{0})\,\, -
\,\,i\,b_{n}(x^{0}) \right]\, e^{- \frac{in\pi } {R} x^{1}} \right\},
\end{eqnarray}
where $ n \epsilon {\cal Z}$. By using (\ref{1}) and (\ref{25}) one finds for
the Lagrangian $ L^{P} $ the expression
\begin{equation}
\label{26}
L^{P}\,\equiv \, \int_{-R}^{+R} dx^{1} {\cal L}\,=\,\frac{1}{2R} \sum_{n>0}
\left[
\omega_{n} (a_{n} \dot{b_{n}}\,-\,\dot{a_{n}} b_{n})\,\,-\,\,\omega_{n}^{2}
(a_{n}^{2}\,+\,b_{n}^{2}) \right].
\end{equation}
Here, $ \omega_{n} \equiv n\pi /R $ and the dot symbolizes derivative with
respect to $x^{0} $. From this Lagrangian follows that the
system possesses a primary first--class constraint
\begin{equation}
\label{27}
p_{a_{0}} \approx 0,
\end{equation}
and a set of primary second--class constraints ($ n > 0 $)
\begin{equation}
\label{28}
\Gamma_{P_{n}}^{\pm}\,\equiv\,p_{a_{n}}\,\pm \,\frac{\omega_{n}}{2R} b_{n}
\approx 0,
\end{equation}
where, $ p_{a_{n}}$ and $ p_{b_{n}} $ are the momenta canonically conjugate to
$ a_{n}$ and $ b_{n}$, respectively. Furthermore, the canonical
Hamiltonian $ H_{c}^{P} $ reads
\begin{equation}
\label{29}
H_{c}^{P}\,=\,\frac{1}{2R} \sum_{n > 0} \omega_{n}^{2}
(a_{n}^{2}\,+\,b_{n}^{2}).
\end{equation}
It is easy to see that there are no secondary constraints. The first--class
constraint $p_{a_{0}}\approx 0 $ is the discrete counterpart of (\ref{22}). It
generates gauge transformations that only affect $ a_{0} $, i.e., the
collective part of $ \phi (x^{0}, x^{1}) $. All phase--space coordinates for $n
> 0 $ are, then, gauge invariant quantities. However, unlike the continuous
case, the gauge freedom can now be suppressed by means of a subsidiary
condition and the system quantized, afterwards, via the Dirac--bracket
procedure. Without loosing generality, we assume for the gauge fixing
condition the functional form
\begin{equation}
\label{30}
a_{0}\,\,+\,\,\xi(a_{n}, b_{n}, p_{a_{n}}, p_{b_{n}}) \approx 0.
\end{equation}
For the $ n > 0 $ sector the nonvanishing commutators turn out to be
\begin{mathletters}
\begin{eqnarray}
\label{31}
&& [\hat{a}_{n}\,\,,\,\,\hat{b}_{m}] \,\, = \,\,-\frac{iR}{\omega_{n}}
\,\delta_{n,m} , \label{mlett:6} \\
&& [\hat{a}_{n}\,\,,\,\,\hat{p}_{a_{m}}] \,\, =  \,\,\frac{i}{2}
\,\delta_{n,m} ,\label{mlett:7} \\
&& [\hat{b}_{n}\,\,,\,\,\hat{p}_{b_{m}}] \,\,  =  \,\,\frac{i}{2}
\delta_{n,m}, \label{mlett:8} \\
&& [\hat{p}_{a_{n}}\,\,,\,\,\hat{p}_{b_{m}}] \,\,  =  \,\,
-\frac{i \omega_{n}}{R} \,\delta_{n,m}. \label{mlett:9}
\end{eqnarray}
\end{mathletters}
As far as the commutators involving $ \hat{a}_{0} $ and/or $ \hat{p}_{a_{0}} $
are concerned, we mention that they can be explicitly computed only after
specifying the gauge function $\xi$ . The Hamiltonian $ H_{c}^{P}$ can be
promoted to the quantum level straightforwardly because is not afflicted by
ordering ambiguities. Then, the solving of the Heisenberg equations of motion
for the independent phase--space variables yields
\begin{equation}
\label{33}
\hat{a}_n(x^0)=\sqrt {\frac{\pi}{2\omega _n}}\,\,\hat{\Lambda}_{n}
e^{-i\omega _n x^0}
+\sqrt {\frac {\pi}{2\omega _n}} \,\,\hat{\Lambda}_{n} ^{\dagger}
e^{i\omega _nx^0},
\end{equation}
where
\begin{equation}
\label{34}
\mbox [\hat{\Lambda}_{n},\hat{\Lambda}_{m}]
= [\hat{\Lambda}_{n}^{\dagger},\hat{\Lambda}_{m}^{\dagger}]=0; \,\,\,
[\hat{\Lambda}_{n}\,\,,\,\,\hat{\Lambda}_{m}^{\dagger}]
= \frac {R}{\pi} \delta_{n,m}.
\end{equation}
The operators $ \hat{\Lambda}_{n}$ and $\hat{\Lambda}_{n}^{\dagger}$
are destruction and creation operators, respectively. Thus, the space of states
is of positive define norm although the vacuum ($|0>$) is not unique. Indeed,
(\ref{25}) can be casted as follows
\begin{equation}
\label{35}
\hat{\phi}(x^{0}\,,\,x^{1})\,=\, - \frac{\hat{\xi}}{2R}\,
+\,\frac{1}{\sqrt{2\pi}}
\left( \frac{\pi}{R} \right) \sum_{n > 0} \frac{1}{\sqrt{\omega _n}}
\left[ \hat{\Lambda}_{n}
e^{-i\omega _n (x^0 + x^{1})}\,+\,\hat{\Lambda}_{n} ^{\dagger} e^{i\omega
_n(x^0+x^{1}) } \right],
\end{equation}
implying that
\begin{equation}
\label{36}
<0|\hat{\phi}(x^{0}\,,\,x^{1})|0>\,=\, - \frac{1}{2R} <0|\hat{\xi}(x^{0})|0>
=  - \frac{1}{2R} <0|\hat{\xi}(0)|0> \neq 0  ,
\end{equation}
where we have used $  \hat{\Lambda}_{n}|0> = 0 $ and the fact that all vacua
are translationally invariant.

The collective mode in (\ref{35}) is neither gauge invariant nor chiral and
is responsible, as just observed, for the spontaneous breaking of the
continuous symmetry $ \phi \rightarrow \phi\,+\,constant $ (see (\ref{1})).
That this remains true at the limit $ R \rightarrow \infty$ follows from
purely dimensional arguments.  Nevertheless, a gauge invariant field operator
($\hat{\Phi}$) with vanishing vacuum expectation value can be naturally built
within the discretized formulation,
\begin{eqnarray}
\label{37}
\hat{\Phi}(x^{0}\,,\,x^{1})\,\,& \equiv & \,\, \hat{\phi}(x^{0}\,,\,x^{1})\,
+\, \frac{\hat{\xi}}
{2R} \nonumber \\
& = & \frac{1}{\sqrt{2\pi}}
\left( \frac{\pi}{R} \right) \sum_{n > 0} \frac{1}{\sqrt{\omega _n}}
\left[ \hat{\Lambda}_{n}
e^{-i\omega _n (x^0 + x^{1})}\,+\,\hat{\Lambda}_{n} ^{\dagger} e^{i\omega
_n(x^0+x^{1}) } \right].
\end{eqnarray}
Clearly, the field $\hat{\Phi}$ only describes chiral excitations. Furthermore,
 it obeys the equal--time commutation relation
\begin{equation}
\label{38}
\mbox{} [\hat{\Phi}(x^{0},x^{1})\,,\,\hat{\Phi}(x^{0},y^{1})] \,
=\,-\frac{i}{2} \left\{
\frac{1}{i\pi} {\sum_{n=-\infty}^{+\infty}}^{\prime} \left(\frac{\pi}{R}
\right)
\frac{e^{\frac{in\pi}{R} (x^{1}-y^{1})}}{\frac{n\pi}{R}} \right\}.
\end{equation}

We investigate next the limit $ R \rightarrow \infty $ for the results obtained
within the discrete approach. In particular, the right hand sides of (\ref{37})
and (\ref{38}) go, respectively, to
\begin{equation}
\label{39}
\hat{\Phi} (x^{0},x^{1}) = \frac{1}{\sqrt{2\pi }} \int _{0^{+}}^{\infty} dk^{1}
\frac{1}{\sqrt{k^{1}}} \left[ \hat{\Lambda}
(k^{1})\,e^{-ik^{1}(x^{0}+x^{1})}\,\,+\,\,\hat{\Lambda}^{\dagger}
(k^{1})\,e^{ik^{1}(x^{0}+x^{1})} \right],
\end{equation}
\begin{equation}
\label{40}
\mbox{} [\hat{\Phi}(x^{0},x^{1})\,,\,\hat{\Phi}(x^{0},y^{1})] \,
=\,-\frac{i}{2} \epsilon
(x^{1}\,-\,y^{1}),
\end{equation}
where the linear momentum variable $ k^{1} $ is the continuous version of the
discrete variable $ n\pi /R $ , while $ \hat{\Lambda}(k^{1}) $ and
$\hat{\Lambda}^{\dagger}(k^{1}) $ are the limiting forms of the corresponding
discrete destruction and creation operators . Of course, as $ R \rightarrow
\infty $ (\ref{34}) maps onto (\ref{10}). Then, the
gauge invariant field $ \hat{\Phi} $ remains chiral at the continuous limit
and, moreover, (\ref{39}) and (\ref{40}) agree with (\ref{9}) and
\ref{mlett:1}, respectively. Hence, what was done in Ref.\cite{Co} by choosing
arbitrarily $\zeta = 0$, amounts to isolate, within the continuous
framework, the gauge invariant piece of the operator $\hat{\phi}$.

As for the Poincar\'e invariance of the present formulation of the FJ model, we
start by recalling that the classical energy--momentum tensor
($ \Theta^{\mu \nu} $), arising from the noncovariant Lagrangian density
(\ref{1}), fulfills $ \partial_{\mu} \Theta^{\mu \nu} = 0 $ but is not
symmetric. In fact, one finds that
\begin{mathletters}
\label{41}
\begin{eqnarray}
\Theta^{0 0}\,=\,-\,\Theta^{0 1}\,=\,\Theta^{1 1}\,=\,\frac{1}{2}\,
(\partial_{1} \phi) (\partial_{1} \phi)\,, \label{mlett:10} \\
\Theta^{1 0}\,=\,\frac{1}{2}\,(\partial_{0} \phi) (\partial_{0}
\phi)\,-\,(\partial_{0} \phi)\,(\partial_{1} \phi). \label{mlett:11}
\end{eqnarray}
\end{mathletters}
The classical components of the energy--momentum tensor serve as a clue for
establishing the form of the quantum densities $ \hat{\Theta}^{\mu \nu} $ in
terms of the basic fields\cite{Sch}. A gauge invariant and symmetric quantum
energy momentum tensor can be constructed by formally replacing $\phi$ by
$\hat{\Phi} $ in (2.20),
\begin{eqnarray}
\label{42}
&&\hat{\Theta}^{0 0}\,=\,-\hat{\Theta}^{0 1}\,=\,-\hat{\Theta}^{1 0}\,=\,
\hat{\Theta}^{1 1}\,= \,  \frac{1}{2}\,: (\partial_{1} \hat{\Phi})
(\partial_{1} \hat{\Phi}) :  \nonumber \\
&& = \frac{1}{4\pi} \left( \frac{\pi}{R} \right)^{2}  \sum_{n>0,m>0}
\sqrt{\omega_{n} \omega_{m}} \left[ -  \hat{\Lambda}_{n} \hat{\Lambda}_{m}
e^{-i(\omega_{n}+\omega_{m})(x^{0}+x^{1})}\,-\,\hat{\Lambda}^{\dagger}_{n}
\hat{\Lambda}^{\dagger}_{m} e^{i(\omega_{n}+\omega_{m})(x^{0}+x^{1})} \right.
\,\,\,\,\,\,\,\,\,\, \mbox{}
\nonumber \\
&& \left. \mbox{} \hskip 5cm  +  2 \hat{\Lambda}^{\dagger}_{n}
\hat{\Lambda}_{m} e^{i(\omega_{n}-
\omega_{m})(x^{0}+x^{1})} \right] ,
\end{eqnarray}
where (\ref{37}) has been used. The normal ordering prescription
introduced in (\ref{42}) secures that $ <0|\hat{\Theta}^{\mu \nu}|0> = 0 $.
We emphasize that the the symmetric character of $ \hat{\Theta}^{\mu \nu} $ is
a consequence of the chiral nature of $\hat{\Phi} $.

The next step consists in investigating the equal--time commutation relations
verified by the quantum energy--momentum densities. In view
of (\ref{42}), there is only one commutator of interest, namely,
$[\hat{\Theta}^{00}(x^{0},x^{1})\,,\, \hat{\Theta}^{00}(x^{0},y^{1})]$. After
some calculations one arrives to
\begin{eqnarray}
\label{43}
\mbox [\hat{\Theta}^{00}(x^{0},x^{1})\,,\, \hat{\Theta}^{00}(x^{0},y^{1})]\,& =
& \,i\,\left(\hat{\Theta}^{00}(x^{0},x^{1})\,+\,\hat{\Theta}^{00}(x^{0},y^{1})
\right) \partial_{x^{1}} \delta(x^{1}\,-\,y^{1}) \nonumber \\
& - &\,i\, \Delta(x^{1}\,,\,y^{1}),
\end{eqnarray}
where
\begin{eqnarray}
\label{44}
&&\Delta(x^{1}\,,\,y^{1})\,=\,\frac{i}{8R^{2}}\sum_{n>0,m>0}
\omega_{n}\,\omega_{m} \left[e^{-i(\omega_{n+m})(x^{1}-y^{1})}\,-\,
e^{+i(\omega_{n+m})(x^{1}-y^{1})} \right] \nonumber \\
&&= \frac{1}{24\pi}\partial_{x^{1}}^{3} \delta(x^{1}\,-\,y^{1})
+\frac{\pi}{24R^{2}} \partial_{x^{1}}\delta(x^{1}\,-\,y^{1})\,\,.
\end{eqnarray}
The algebra (\ref{43}) is a Virasoro algebra, since the additional piece in
the right hand side of (\ref{44}) ($(\pi/24R^{2})
\partial_{x^{1}}\delta(x^{1}\,-\,y^{1}) $) is a trivial cocycle that can be
absorved in a constant redefinition of the energy--momentum tensor.

We now show that the charges arising from $\Theta^{\mu \nu}$ are the
generators of the Poincar\'e symmetry.
Indeed, by integrating both sides of (\ref{43}) over the domain of the variable
   $y^{1}$
and after taking into account that
\begin{equation}
\label{45}
\int_{-R}^{+R} dy^{1} \Delta(x^{1}\,,\,y^{1})\,=\,0 ,
\end{equation}
one obtains
\begin{equation}
\label{46}
\mbox [\hat{\Theta}^{00}(x^{0},x^{1})\,,\, \hat{P}^{0}]\,=\,i \partial_{x^{1}}
\hat{\Theta}^{00}(x^{0},x^{1}),
\end{equation}
where
\begin{equation}
\label{47}
\hat{P^{0}}\,\equiv \,\int_{-R}^{+R}dx^{1}
\hat{\Theta}^{00}(x^{0},x^{1})\,=\,\frac{\pi}{R} \sum_{n>0} \omega_{n}
\hat{\Lambda}^{\dagger}_{n}\hat{\Lambda}_{n}\,=\,:\hat{H}^{P}_{c}:
\end{equation}
is the generator of translations in time. A subsequent $ x^{1} $ integration in
(\ref{46}), including the factor $ x^{1} $, gives
\begin{equation}
\label{48}
\mbox [\hat{P^{0}}\,,\,\hat{M}]\,=\,i\,\hat{P^{0}},
\end{equation}
where
\begin{equation}
\label{49}
M \,\equiv\, -\,x^{0}\,
\hat{P}^{1}\,+\,\int_{-R}^{+R}dx^{1}x^{1}\,\hat{\Theta}^{00}(x^{0},x^{1})
\end{equation}
is the Lorentz boosts generator and $\hat{P^{1}}=-\hat{P^{0}}$ is the
generator of spatial translations. What we have in (\ref{48}) is, precisely,
the contracted Poincar\'e algebra of Ref. \cite{Fl}.
This concludes our study of the FJ model on a compact domain and under periodic
boundary conditions.

We turn next into investigating the FJ model under antiperiodic boundary
conditions.
In this case, the transformation (\ref{21}) is not allowed because it does not
preserve the boundary conditions \cite{Bg}. Hence, the theory exhibits no
symmetry and, correspondingly, first-class constraints should not arise.
In other words, under antiperiodic boundary conditions, the FJ model is a pure
second-class system whose excitations are all chiral. To check that this is
indeed the case, we go again into the discrete formulation.

Instead of (\ref{25}) we write, in the present case,
\begin{equation}
\label{50}
\phi(x^{0}\,,\,x^{1}) = \frac{1}{2R} \sum_{n \geq 0}\left\{
\left[ a_{n}(x^{0})+ib_{n}(x^{0}) \right]
e^{i \omega_{n+\frac{1}{2}} x^{1}}+\left[ a_{n}(x^{0})-ib_{n}(x^{0})
\right] e^{-i \omega_{n+\frac{1}{2}} x^{1}} \right\},
\end{equation}
which together with (\ref{1}) leads to the Lagrangian
\begin{equation}
\label{51}
L^{A}\,=\,\frac{1}{2R} \sum_{n \geq 0} \left[\omega_{n+\frac{1}{2}}
(a_{n} \dot{b_{n}}\,-\,\dot{a_{n}} b_{n})\,\,-\,\,
\omega_{n+\frac{1}{2}}^{2}
(a_{n}^{2}\,+\,b_{n}^{2}) \right].
\end{equation}
It is now easy to convince oneself that {\it all} the constraints
($n \geq 0$),
\begin{equation}
\label{52}
\Gamma_{A_{n}}^{\pm}\,\equiv\,p_{a_{n}}\,\pm \,
\frac{\omega_{n + \frac{1}{2}}}{2R} b_{n} \approx 0,
\end{equation}
are, in fact, second class.

The system is quantized by using, once more, the Dirac-bracket quantization
procedure. We omit the details and just mention that, this time, the quantum
dynamics is solved by the field operator
\begin{equation}
\label{53}
\hat{\phi}(x^{0}\,,\,x^{1})\,=\, \frac{1}{\sqrt{2\pi}}
\left( \frac{\pi}{R} \right) \sum_{n \geq 0}
\frac{1}{\sqrt{\omega_{n+\frac{1}{2}}}}
\left[ \hat{\Lambda}_{n}
e^{-i\omega_{n+\frac{1}{2}} (x^0 + x^{1})}\,+\,\hat{\Lambda}_{n} ^{\dagger}
e^{i\omega_{n+\frac{1}{2}}(x^0+x^{1}) } \right],
\end{equation}
where $\hat{\Lambda}_{n}$ and $\hat{\Lambda}_{n}^{\dagger}$ are, for all values
of $ n $, destruction and creation operators, respectively.

Therefore, the vacuum expectation value of $\hat{\phi}$ vanishes implying that,
for antiperiodic boundary conditions, the vacuum is unique. We also remark that
the chiral nature of $\hat{\phi}$ survives the limit $R \rightarrow \infty$ and
that (\ref{mlett:1}) is obeyed.

Since the field $\hat{\phi}$ is itself chiral, a symmetric quantum energy-
momentum tensor can be constructed by using the first equality of (\ref{42})
with  $\hat{\Phi}$ replaced by  $\hat{\phi}$, namely,
\begin{equation}
\label{54}
\hat{\Theta}^{0 0}\,=\,-\hat{\Theta}^{0 1}\,=\,-\hat{\Theta}^{1 0}\,=\,
\hat{\Theta}^{1 1}\,= \,  \frac{1}{2}\,: (\partial_{1} \hat{\phi})
(\partial_{1} \hat{\phi}) : \,\,.
\end{equation}
As for periodic boundary conditions, the equal--time algebra of densities
is a Virasoro algebra. Moreover, a set of charges can be
constructed which act as generators of the Poincar\'e group.

To summarize, after recognizing the improper nature of the constraints, we
were able of
performing a consistent quantization of the FJ model as a constrained system.
It became clear that each boundary condition defines a different problem,
mainly in connection with the symmetry content of the theory: while the
periodic case is a gauge theory, the antiperiodic is not. In both cases the
physical excitations are all chiral.


\begin{references}
\bibitem {F1} Throughout this paper we use natural units $ (c = \hbar =
1)$. Our metric is $ g_{00} = -g_{11} = 1 $.
\end{references}
\end{document}